\documentclass[aps, prl,superscriptaddress, twocolumn,showpacs,preprintnumbers,amsmath,amssymb]{revtex4}
\usepackage{graphicx}
\usepackage{bm}
\usepackage{amssymb, latexsym, amsmath, textcomp}

\renewcommand{\v}[1]{{\bf #1}}

\newcommand{\s}{{\sigma}}

\def\be{\begin{eqnarray}}
\def\ee{\end{eqnarray}}

\newcommand{\Eq}[1]{Eq.~(\ref{#1})}

\newcommand{\ua}{\uparrow}
\newcommand{\da}{\downarrow}

\begin{document}
\title{Spin-charge Separated Solitons in a Topological Band Insulator}
\author{Ying Ran}
\author{Ashvin Vishwanath}
\author{Dung-Hai Lee}
\affiliation{Department of Physics, University of California at
Berkeley, Berkeley, CA 94720, USA} \affiliation{Material Science
Division, Lawrence Berkeley National Laboratory, Berkeley, CA 94720,
USA}
\date{\today}

\begin{abstract}
In this paper we construct a simple, controllable, two dimensional
model based on a topological band insulator. It has many attractive properties. 
(1) We obtain spin-charge separated solitons that are associated
with $\pi$ fluxes. (2) It suggests an alternative way to classify
$Z_2$ topological band insulator without resorting to the
sample boundary. 
 (3) When the $\pi$ fluxes are dynamical variables, as in a correlated insulator with emergent gauge fluxes,
 these solitons are propagating bosonic excitations and their condensation triggers a phase transition into a planar ferromagnet.
\end{abstract}
\maketitle

%Spin charge separated quasiparticles in more than one space
%dimension ($d$) is of tremendous interest in condensed matter
%physics. Unlike in $d=1$, this phenomenon in $d\ge 2$ signifies the
%existence of unusual, usually topologically ordered,\cite{wen}
%ground state. For $d>1$ well controlled models exhibiting spin
%charge separation, if exist, are extremely rare. In this paper we
%discuss such a model. As byproducts, our model also leads to  a
%number of other results that are connected to other topics of great
%current interest in condensed matter physics.

There has been recent interest in novel varieties of band insulators
which differ in subtle but essential ways from ordinary band
insulators. The best known example is the Chern insulator
\cite{tkkn} which breaks time reversal symmetry. Tight biding models
of such insulator has been studied long back by Hofstadter. More
recently, a tight binding for Chern insulator which has no net
magnetic flux was proposed by Haldane.\cite{haldane} %on the honeycomb
%lattice.
Recent progress has focused on time reversal invariant
insulators, where a natural generalization of the Haldane model
emerges on including spin orbit interactions \cite{km}. These models
have attracted considerable attention recently because of their
relevance to the quantum spin Hall effect.\cite{spinhall} Their band
structures are characterized by non-trivial $Z_2$ topological
quantum number\cite{km,moore}, which differentiates them from
ordinary band insulators, hence the terminology topological band
insulators (TBIs). Here we point out a remarkable property of these
insulators - the presence of spin charge separated solitons in the
presence of $\pi$ flux - which allows for a bulk definition of the
TBI. Spin charge separated excitations
%quasiparticles in more than one space
%dimension ($d$) is
are of tremendous interest in condensed matter physics. While
spin-charge separation is common in 1D, in higher dimensions it is
extremely rare, requiring the presence of novel quantum states with
topological order \cite{wen}. The spin-charge separated solitons
identified here are not propagating excitations since they are tied
to external $\pi$ flux. However, if the flux itself is a dynamical
variable, then we show that these solitons are bosonic excitations
and different type of solitons obey mutual semionic statistics.

{\bf The model}~We will start from a topological band insulator
(TBI) fermion model, and then couple it with a dynamical $Z_2$ gauge
field. TBIs are free fermion insulators. Their band structures are
characterized by non-trivial topological quantum
number\cite{tkkn,km,moore}, which differentiates them from ordinary
band insulators. Two well known examples of TBI are the models
proposed by Haldane\cite{haldane} (time-reversal breaking) and the
$Z_2$ insulator model proposed by Kane and Mele (time reversal
invariant)\cite{km} on the honeycomb lattice. %These models have
%attracted considerable attention recently because of their possible
%relevance to quantum spin Hall effect.\cite{spinhall}
The Hamiltonian of our TBI, which is a square lattice version of the
Kane-Mele model, is given by the Hamiltonian below with the bond
variables $\tau^z_{ij}$ set to $+1$ everywhere:
%is given by the first line of \Eq{model_ham} with
%$\tau^z_{ij} =+1$ everywhere,
\begin{widetext}
\begin{align}
H_0=&\sum_{i,\sigma}\Big\{\Psi_{i,\sigma}^{\dagger}\begin{pmatrix}
                                -\nu/2  &0\\
                                 0      &\nu/2
                               \end{pmatrix} \Psi_{i,\sigma}
                               +%\sum_{i\sigma}
                               %\left[
                               \tau^z_{i,i+x}\Psi^{\dagger}_{i+x,\sigma}\begin{pmatrix}
                                       -t     & -\Delta\\
                                       \Delta &  t
                                       \end{pmatrix}\Psi_{i,\sigma} + \tau^z_{i,i+y}\Psi_{i+y,\sigma}^{
\dagger}\begin{pmatrix}
                                            -t & \mathrm{i}~{\rm sign}(\sigma)\Delta\\
                                          \mathrm{i}~{\rm sign}(\sigma)\Delta & t
                                           \end{pmatrix}\Psi_{i,\sigma}%\right]
                                           \Big\}+h.c.
                                           %\notag
                                           %\\&-K\sum_{\square}\big(\prod_{\langle
%ij\rangle\in\square}\tau_{ij}^z\big)+J\sum_{\langle ij
%\rangle}(\tau^x_{ij}).
\label{model_ham}
\end{align}
\end{widetext}
Here $i$ labels the sites of the square lattice,
$\Psi^\dagger_\s=(c^\dagger_{1\s},c^\dagger_{2\s})$ is a
two-component electron operator where $\sigma$ is the spin index and
$1,2$ are the flavor (or `orbital') indices. These play the same
role as sublattice indices in the honeycomb model. Also, $t,\Delta$
are real hopping parameters while $\nu$ is an onsite orbital
splitting energy. This free-fermion Hamiltonian is invariant under
the z-axis spin rotation ($R_{S_z}$). Moreover, since the hopping
matrices of the up spin fermions are the hermitian conjugate of that
of the down spin fermions, it is also invariant under time-reversal
(T). So long as $\Delta\ne 0$ this model is in the TBI phase when
$0<|\nu|<8|t|$.  Note that if one ignores the spin down fermions and
replaces $c_2$ by $c_1^{\dagger}$ for the up spin fermion, the free
fermion Hamiltonian discussed above becomes the Bogoliubov
Hamiltonian of a spin polarized p+ip superconductor\cite{majorana}.
The main difference between our free-fermion Hamiltonian and that of
the p+ip superconductor is symmetry : while charge is conserved in
the former (hence possess global $U(1)$ symmetry), it is only
conserved modulo two (hence possess global $Z_2$ symmetry only) in
the latter. It is well known that the vortex cores of the $p+ip$
superconductor has Majorana fermion zero modes\cite{majorana}. As
shown in Ref.\cite{bernevig} the corresponding $U(1)$ symmetric
model describes a TBI and possess a pair of opposite spin,
counter-propagating, edge states at the sample boundary, analogous
to the Kane-Mele model\cite{km}.

In Ref.\cite{lzx} it was shown that when the hopping between the
edges is reinserted with a twist in sign, equivalent to flip the
sign of half a row of $\tau_{i,i+y}$ in \Eq{model_ham} and hence
introducing a $\pi$ flux, an edge Jackiw-Rebbi soliton is
created\cite{lzx}. Such a soliton is a {\it point defect in
two-dimensions}, which possesses two fermionic (not Majorana
fermion) zero modes, one for each spin\cite{lzx}. We call these
soliton defects {\em ``fluxons''}.

%We have introduced the bond variables $\tau^z_{ij} =\pm 1$ to allow
%for $\pi$-flux plaquettes in the band structure. Later, we will make
%these  fluxes dynamical variables. Moreover, when the hopping
%between the edges is reinserted with a twist in sign, {\bf
%equivalent to introducing $\pi$ flux,} an edge Jackiw-Rebbi soliton
%is created\cite{lzx}. Such a soliton is a {\it point defect in
%two-dimensions}, which possesses two fermionic (not Majorana
%fermion) zero modes, one for each spin\cite{lzx}. We call these
%defects {\em ``fluxons''}.
% we shall discuss in the rest of the paper are
%precisely this type of defects.

{\bf Spin-charge separated fluxons} %To study the effect of
%externally defined static $\pi$ fluxes on the TBI, we introduce bond
%variables $\tau^z_{ij} =\pm1$.
If in a plaquette
%In the following we
%shall refer to $\prod_{ij\in\square}\tau_{ij}^z=+1$ as no flux and
$\prod_{ij\in\square}\tau_{ij}^z=-1$, this signifies a fluxon. We
have performed numerical calculations and  shown that for a wide
range of the hopping parameters $\nu,t,\Delta$ the creation energy
of two fluxons are lower than the minimum energy for particle-hole
excitations, i.e., the band gap. As shown in Ref.\cite{lzx}, there
are two fermionic midgap modes localized on each fluxon, which are
Kramers conjugate. Since the model (\ref{model_ham}) enjoys two
independent particle hole symmetries $PH_{\s}:c_{1\s}\rightarrow
c_{2\s}^{\dagger},c_{2\s}\rightarrow c_{1\s}^{\dagger}$ where
$\s=\uparrow$ or $\s=\downarrow$, these modes are precisely at zero
energy (i.e. in the middle of the gap).

\begin{figure}
\includegraphics[scale=0.5]{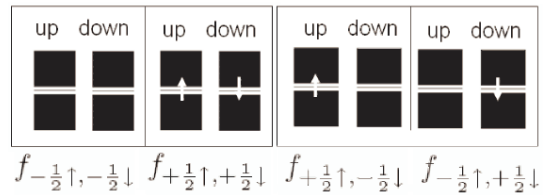}
\caption{Four different types of fluxons. The notation, e.g.,
$f_{+\frac{1}{2}\uparrow,-\frac{1}{2}\downarrow}$ means the up-spin
zero-mode is filled while down-spin zero-mode is empty.
\label{fluxon}}
\end{figure}

\begin{figure}
 \includegraphics[width=0.21\textwidth]{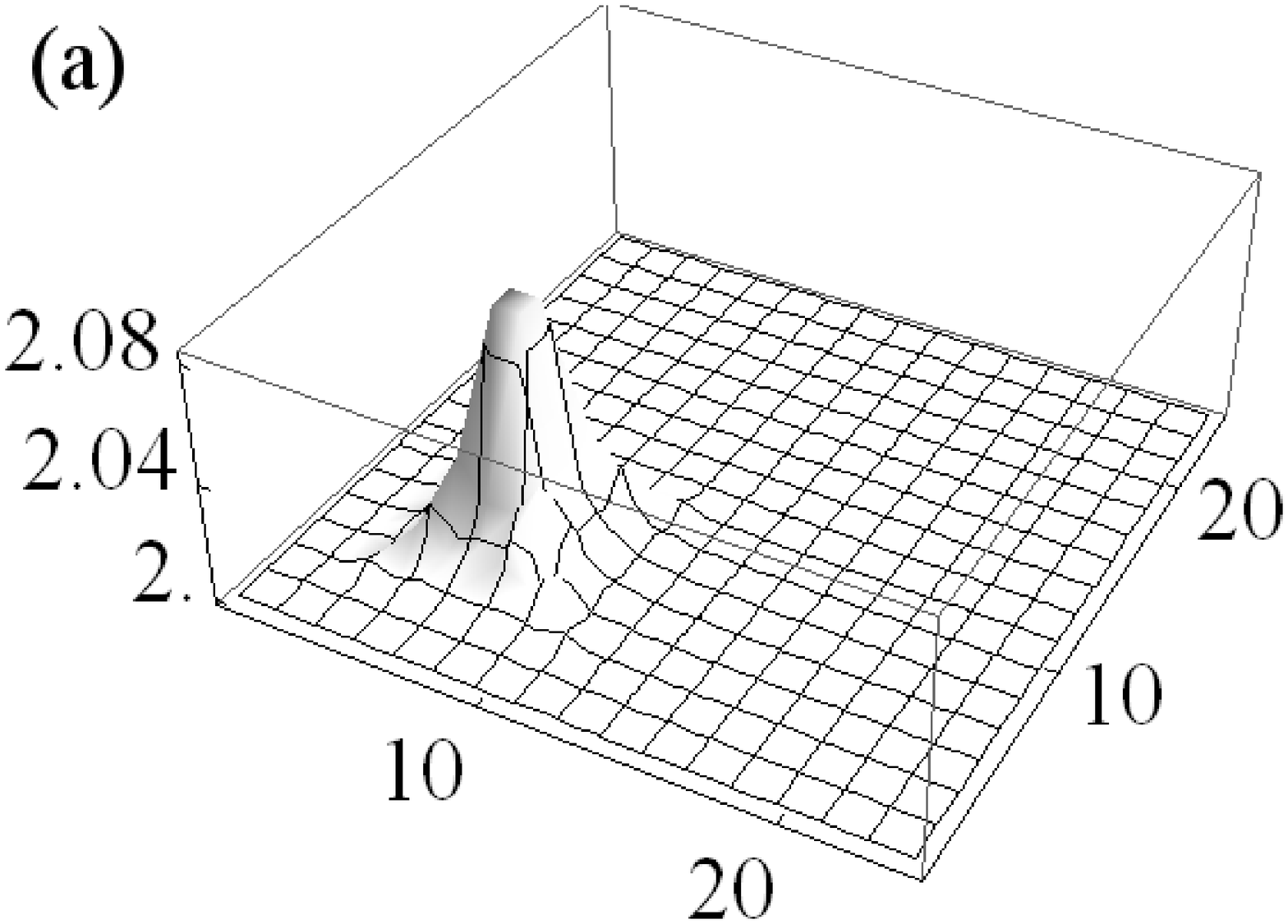}
 \;\;\includegraphics[width=0.21\textwidth]{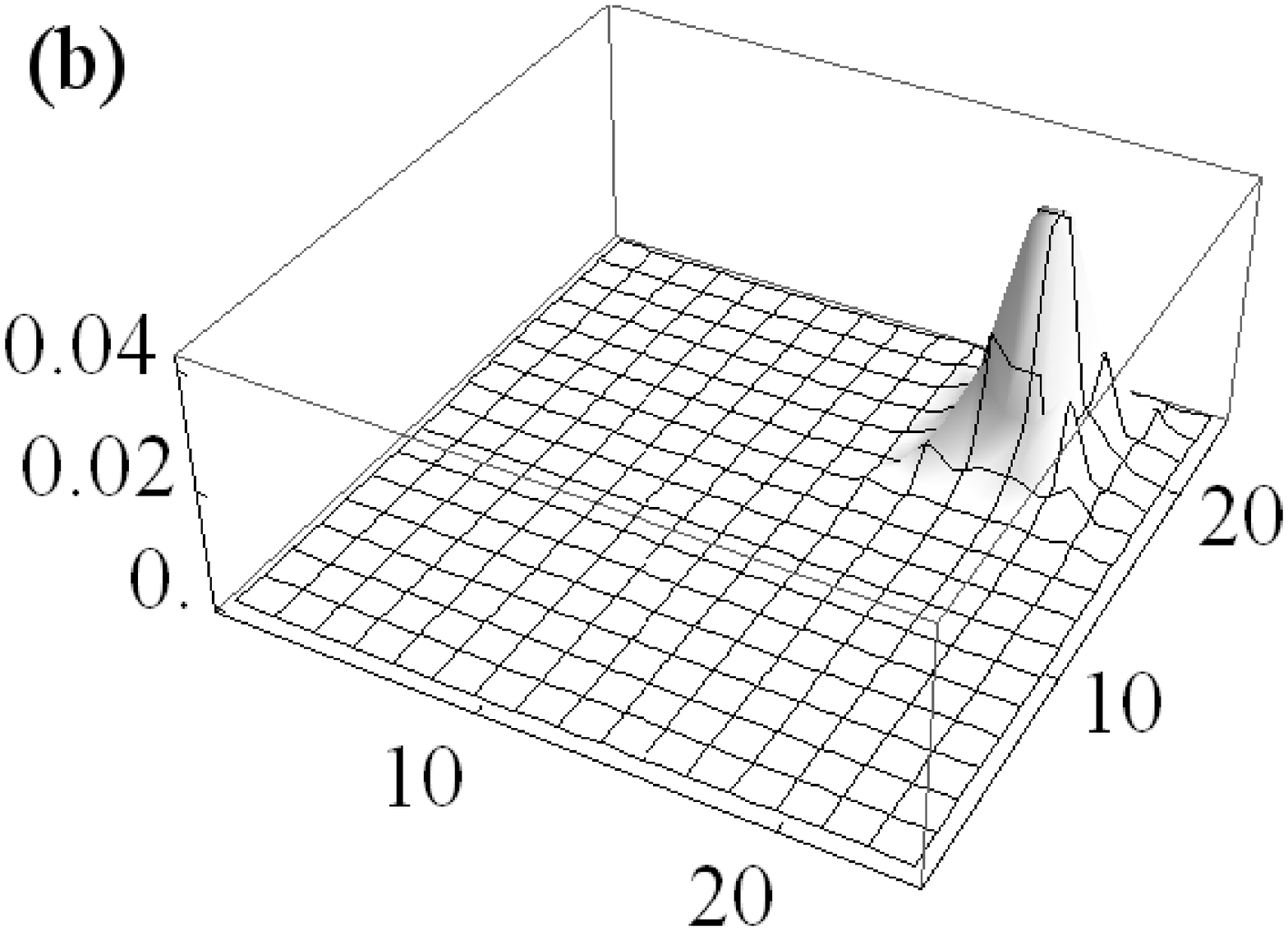}
\caption{The charge (a) and spin (b) densities of a pair of fluxons
on a $24\times 24$ lattice with periodic boundary condition. The
fluxon at coordinate $(6,6)$ is $f_{+{1\over 2}\ua,+{1\over 2}\da}$.
It has charge $1$ and $S_z=0$. The fluxon at $(18,18)$ is
$f_{+{1\over 2}\ua,-{1\over 2}\da}$, it has charge $0$ and
$S_z=1/2$. Note that that the average charge density of the ground
state is $2$. The spin-charge separation in real space is manifest.}
\label{spin-charge}
\end{figure}

The occupation/unoccupation of these zero modes leads to an
excess/deficit of 1/2 fermion number per spin. The four different
ways of occupying these zero modes (Fig.(\ref{fluxon})) give rise to
four different types of fluxons with the following quantum numbers:
$f_{+{1\over 2}\ua,+{1\over 2}\da}$ (charge 1, $S_z=0$);
$f_{-{1\over 2}\ua,-{1\over 2}\da}$ (charge -1, $S_z=0$);
$f_{+{1\over 2}\ua,-{1\over 2}\da}$ (charge 0, $S_z={1\over 2}$);
$f_{-{1\over 2}\ua,+{1\over 2}\da}$ (charge 0, $S_z=-{1\over 2}$).
The presence of these modes, and their quantum numbers, can also be
deduced from flux threading arguments for the up and down spin
integer quantum Hall states. In the absence of particle-hole
symmetry the modes are no longer precisely at zero energy, but must
still be within the gap. As discussed subsequently, this structure
is essentially preserved even when spin rotation symmetry is
completely broken, as long as time reversal symmetry remains.
 In Fig.\ref{spin-charge} we present the charge and spin
density profiles for a pair of $f_{+{1\over 2}\ua,+{1\over 2}\da}$
and $f_{+{1\over 2}\ua,-{1\over 2}\da}$ fluxon.

\begin{figure}
\includegraphics[scale=0.4]{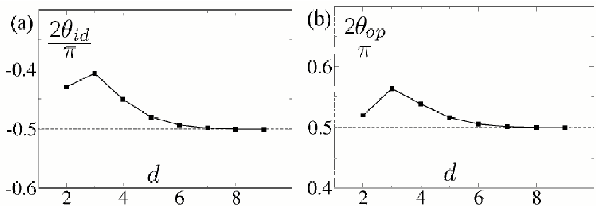}
\caption{The result for $\theta(f_{1\ua},f_{2\ua})$ on a $4d\times
4d$ square lattice with periodic boundary condition. We fix the
position of $f_{2\ua}$ and let $f_{1\ua}$ loops around along a
$2d\times 2d$ square path. The fermionic band parameters used in the
computation are: $\nu=0.3,\Delta=0.5,t=1$.(a) Identical fluxons
$\theta_{id}\equiv
\theta(f_{\frac{1}{2}\uparrow},f_{\frac{1}{2}\uparrow})
=\theta(f_{-\frac{1}{2}\uparrow},f_{-\frac{1}{2}\uparrow})$, and (b)
distinguishable fluxons
$\theta_{op}\equiv\theta(f_{-\frac{1}{2}\uparrow},f_{\frac{1}{2}\uparrow})
=\theta(f_{\frac{1}{2}\uparrow},f_{-\frac{1}{2}\uparrow})$. Note
that due to the particle-hole symmetry
$\theta(f_{\frac{1}{2}\uparrow},f_{\frac{1}{2}\uparrow})
=\theta(f_{-\frac{1}{2}\uparrow},f_{-\frac{1}{2}\uparrow})$ and
$\theta(f_{-\frac{1}{2}\uparrow},f_{\frac{1}{2}\uparrow})=\theta(f_{\frac{1}{2}\uparrow},f_{-\frac{1}{2}\uparrow})$.
 The extrapolation to
$d\rightarrow\infty$ gives $-\theta_{id}=\theta_{op}=\pi/4$.
\label{num_stat}}
\end{figure}

{\bf Quantum statistics of fluxons} When fluxons are mobile
%($J\ne 0$)
 their quantum statistics becomes important. %{\bf I think we can
%define statistics even if the fluxons are not dynamical - we can hop
%them around in a closed loop by adiabatic variations of the
%Hamiltonian}
We determine their statistics through explicit computation of the
Berry's phase. The statistical angle between two fluxons (not
necessarily identical), $\theta(f_1,f_2)$, is defined as 1/2 times
the difference of the following two Berry's phases. The first is
obtained by hopping $f_1$ in a clockwise loop enclosing $f_2$, while
the second is obtained by hopping $f_1$ along the same path but with
$f_2$ sitting outside the loop. Given a closed loop sequence of
fluxon positions $\{(\v x_1^i,\v x_2^i),~~i=0,1,2,\cdots N \}$ the
Berry's phase is given by
\begin{align}
 \theta=\mbox{Im}\ln\left[\prod_{i=1}^{N}\langle\Phi(\v x_1^{i+1},\v x_2^{i+1})\vert H_{hop}
 \vert\Phi(\v x_1^{i},\v x_2^{i})\rangle\right],\label{lattice_berry_phase}
\end{align}
where $H_{hop}=J\sum_{i}(\tau^x_{i,x}+\tau^x_{i,y})$, and
$\vert\Phi(\v x_1,\v x_2)\rangle$ is the fermion many-body ground
state consistent with two fluxons being at $\v x_1$ and $\v
x_2$\cite{note}. Since the up-spin band and the down-spin band decouples, the whole electronic wavefunction is a product of two Slater-determinants $\vert\Phi(\v x_1,\v x_2)\rangle=\vert\Phi(\v
x_1,\v x_2)\rangle_{\ua}\otimes \vert\Phi(\v x_1,\v
x_2)\rangle_{\da}$. As a result, \be
\theta(f_{\alpha_1\uparrow,\beta_1\downarrow},f_{\alpha_2\uparrow,\beta_2\downarrow})=\theta(f_{\alpha_1\ua},f_{\alpha_2\ua})+\theta(f_{\beta_1\da},f_{\beta_2\da}),\label{sum}\ee
where $\theta(f_{\alpha_1\ua},f_{\alpha_2\ua})$ is the statistical angle between the two fluxons in the up-spin band. The results for $\theta(f_{\alpha_1\ua},f_{\alpha_2\ua})$ are presented in
Fig.(\ref{num_stat}). 
They are consistent with the statistics obtained from anyon fusion
arguments\cite{franz}: Let us discuss on the up-spin band only. Consider a bound state of two of fluxons
$f_{\frac{1}{2}\ua}$, and another bound state of two fluxons
$f_{\frac{1}{2}\ua}$. Then each bound state carries charge 1 and flux
$2\pi\sim 0$ and thus is a fermion. As a result the statistical
phase between two $f_{\frac{1}{2}\ua}$ fluxons would be one-quarter of that
of fermions, i.e. $\pm\frac{\pi}{4}$. By numerical calculation we find $\theta(f_{\frac{1}{2}\ua},f_{\frac{1}{2}\ua})=-\frac{\pi}{4}$. From particle-hole symmetry we immediately conclude that $\theta(f_{-\frac{1}{2}\ua},f_{-\frac{1}{2}\ua})=-\frac{\pi}{4}$, too. Now consider a bound state of an $f_{\frac{1}{2}\ua}$ fluxon and an $f_{-\frac{1}{2}\ua}$ fluxon. This bound state carries charge $0$ and should be a boson. This implies that $\theta(f_{\frac{1}{2}\ua},f_{-\frac{1}{2}\ua})=\frac{\pi}{4}$.
$\theta(f_{\alpha_1\ua},f_{\alpha_2\ua})$ one can determine the statistical phase in the down-spin band
$\theta(f_{\alpha_1\da},f_{\alpha_2\da})$ readily: \be
\theta(f_{\alpha_1\da},f_{\alpha_2\da})=-\theta(f_{\alpha_1\ua},f_{\alpha_2\ua}).\label{rel}\ee
This is because the Hamiltonian for the down spin band is the
hermitian conjugate of that for the up spin band. Given
Fig.(\ref{num_stat}) and Eqs.(\ref{sum},\ref{rel}) we have
determined the quantum statistics of fluxons.  The
result is shown in
the following Table. %Basically the fluxons are bosons, and chargeless-fluxon
%and chargeful-fluxon satisfy mutual semion statistics due to the
%Aharonov-Bohm effect.
In general fluxons should experience a background magnetic field
(the fermion density) as they hop around. However, since there are
on average two fermions per site (see Fig.\ref{spin-charge}(a)),
this background magnetic flux is $2\pi$ per plaquette, hence is
equivalent to no flux. %Thus there is no background flux to impede
%the fluxon motion.
The above results should be robust against perturbations so long as
the bulk gap is preserved.
\begin{center}
\begin{tabular}
{|l|l|} \hline
~~~~~~Self statistics& ~~~~~~Mutual statistics\\
\hline ~~~~~~~~$c=\pm 1/2$&~~~~~~~~~~$c=\pm 1/2$\\
\hline$\theta(f_{c\ua,c\da},f_{c\ua,c\da})=~~0$~~ &
\begin{tabular}{|l|}
$\theta(f_{c\ua,c\da},f_{-c\ua,-c\da})=0~~~~~~$\\
\hline$\theta(f_{c\ua,c\da},f_{c\ua,-c\da})={\pi\over 2}$\\
\hline $\theta(f_{c\ua,c\da},f_{-c\ua,c\da})={\pi\over 2}$
\end{tabular}
\\
\hline $\theta(f_{c\ua,-c\da},f_{c\ua,-c\da})=0$ &
\begin{tabular}{|l|} $\theta(f_{c\ua,-c\da},f_{-c\ua,c\da})=0~~~~~~$\\
\hline $\theta(f_{c\ua,-c\da},f_{c\ua,c\da})={\pi\over 2}$\\
\hline $\theta(f_{c\ua,-c\da},f_{-c\ua,-c\da})={\pi\over
2}$\end{tabular}\\
\hline
\end{tabular} \label{st}
\end{center}

{\bf A new way to diagnose $Z_2$ TBI} Note that the four fluxon
states in Fig.\ref{fluxon} are degenerate due to $T,PH_{\ua,\da}.$
The degeneracy between the charged and neutral fluxon can be easily
removed by adding a weak short range charge repulsion to the
original fermion model. After that, one expects the lowest energy
fluxons to be the neutral ones:
$f_{+\frac{1}{2}\uparrow,-\frac{1}{2}\downarrow}$ and
$f_{-\frac{1}{2}\uparrow,+\frac{1}{2}\downarrow}$. In the rest of
the paper we refer to them as spin fluxons. The $S_z=\pm 1/2$ spin
fluxons form a Kramer's pair upon time reversal.

So far in our discussion $R_{S_z}$ is a symmetry of the Hamiltonian.
This global $U(1)$ symmetry justifies the corresponding TBI to be
called a $U(1)$ TBI. However, the presence of a Kramer pair of
neutral fluxon is more general. We have checked that as long as $T$
is unbroken, each neutral fluxon always comes as a Kramer pair. This
is true even after breaking $R_{S_z}$(by adding, say, T-invariant
spin-flip hopping term to the TBI Hamiltonian\cite{km}), and/or
$PH_{\ua\da}$ (by adding, say, a chemical potential term to the TBI
Hamiltonian). This robust degeneracy allows one to diagnose the
$T$-invariant TBI, or $Z_2$ TBI\cite{km} {\it without resorting to
edge states.} For example, consider a $Z_2$ TBI on a torus. One can
introduce $2N$ far apart, low-energy, spin fluxons by, e.g.,
imposing an energy penalty for charge accumulation. The ground state
will be $2^{2N}$-fold degenerate. On the other hand a trivial band
insulator has no such degeneracy. Hence this degeneracy
differentiates a $Z_2$ TBI from a trivial band insulator. This can
be implemented as a numerical diagnosis of $Z_2$ TBIs. %{\bf What if
%the spin degenerate states are higher in energy than the charge
%states? We need to impose a constant charge conditions I think}

This study naturally generalizes to three dimension. For the
3D-$Z_2$ insulator(which is refered as the strong topological
insulator in literatures, for instance \cite{Fu_Kane_Mele,moore}),
we find for a closed $\pi$-flux loop, there are two gapless
one-dimensional Dirac fermion modes propagating along the $\pi$-flux
loop in opposite directions and are Kramers conjugates of each
other.

{\bf Dynamical $\pi$-fluxes and TBI*~} In order to make the fluxon
elementary excitations, we give the $Z_2$ variable dynamics. This is
achieved by adding the following term to \Eq{model_ham}.

\begin{equation}
H_{{\rm TBI}^*}=H_0 -K\sum_{\square}\prod_{\langle
ij\rangle\in\square}\tau_{ij}^z+J\sum_{\langle
ij\rangle}\tau^x_{ij}. \label{model_ham_TBI*}
\end{equation}
The fermions $\Psi_{i,\sigma}$ in the above Hamiltonian carries a
$Z_2$ gauge charge, hence are not ordinary electrons. We refer to
such a correlated band insulator with emergent $Z_2$ gauge fields as
a TBI*. Nonetheless, the fundamental fermion degrees of freedom of
\Eq{model_ham_TBI*} possesses both the fermion number and the spin
quantum number. In the following we show that the elementary
excitations of this model exhibit separation of the the fermion
quantum number (which we abbreviated by ``chanrge'') and spin.
%The Hilbert space is the direct product of those of fermion and
%gauge field, with any two gauge field states related by a $Z_2$
%gauge transformation identified.

In \Eq{model_ham_TBI*} the term $\prod_{\langle
ij\rangle\in\square}\tau_{ij}^z$ is the $Z_2$ gauge flux going
through a plaquette,  and $K,J$ are gauge couplings. The last term
of Eq.(\ref{model_ham_TBI*}) causes the fluxons to hop from one to a
neighboring plaquette. As usual, $\tau^{x,z}$ are the first and
third components of the Pauli matrices. The Hamiltonian in
Eq.(\ref{model_ham_TBI*}) has to be supplemented with a local
constraint on every site (the `Gauss Law') $ \prod_{j \in {\rm n.n.
of~}i}\tau_{ij}^x=(-1)^{\Psi_{i,\sigma}^{\dagger} \Psi_{i,\sigma}}$, where
the product is over nearest neighbors of the site `i'. For $J=0$ the
ground state of Eq.(\ref{model_ham}) lies in the gauge sector where
there is no flux in any plaquette. Under that condition it is
always possible to tune the parameters so that
the fluxons are the lowest energy excitations in the fermionic sector.
%In addition to the fluxons, there
%are also fermion particle-hole excitations. The minimum energy for
%creating these excitations is
%the band gap $E_G$. %Therefore it is important to know whether the
%fluxon excitation lies below or above the particle-hole continuum.
%We have performed numerical calculations and  shown that for a wide
%range of the hopping parameters $\nu,t,\Delta$ the fluxons are lower
%in energy than the particle-hole excitations.
%In addition, due to
%the existence of the parameter $K$, it is always possible to make
%$E_\pi+K<<E_G$ by choosing $K$ to be negative.
For non-zero $J$ the static fluxons are no longer eigen excitations.
However, so long as the fluxon creation energy $>>J$ the
delocalization of fluxons will not close the excitation gap. In that
limit the gapped mobile fluxons exhibit spin-charge separation as
illustrated in Fig.\ref{fluxon}.
%Thus we have achieved in constructing a
%controllable model whose elementary excitations are spin-charge
%separated.

%As written, \Eq{model_ham} is invariant under global $T, R_{S_z}$,
%and two independent particle-hole transformations
%$PH_{\s}:c_{1\s}\rightarrow c_{2\s}^{\dagger},c_{2\s}\rightarrow
%c_{1\s}^{\dagger}$ where $\s=\pm 1$.

{\bf Spin fluxon condensation} In the rest of this paper we will
assume the spin fluxons to be the lowest energy excitations. Now let
us ask what happens as the magnitude of $J$ is increased. When the
energy cost in creating a static spin fluxon is counter balanced by
the kinetic energy gain due to its delocalization, spin fluxons will
spontaneously proliferate. Owing to their Bose statistics this will
trigger Bose condensation at zero temperature. It is interesting to
ask what is the nature of the new ground state and what is the
nature of the (quantum) phase transition. In the following we shall
discuss two scenarios. %The first is when both $S_z$ and T are
%conserved as in Eq.(\ref{model_ham}). The second case is when
%perturbations (such as the ones discussed earlier) have been
%introduced to break the $R_{S_z}$ while maintaining T.

(I) If $S_z$ is conserved, two spin fluxons of opposite $S_z$ can be
created and annihilated dynamically, while two fluxons with the same
$S_z$ can not. In this case we can view the $S_z=-\frac{1}{2}$
fluxon as the anti-particle of $S_z=+\frac{1}{2}$ fluxon, and  T
transforms one into the other. The symmetry which dictates the $S_z$
conservation is $R_{S_z}$.  Under such condition, the field theory
describing the spin fluxon condensation is characterized by the
following Lagrangian density
\begin{align}
 \mathcal{L}=\frac{1}{2}\vert\partial_{\tau}\phi\vert^2
 +\frac{1}{2}\vert\nabla\phi\vert^2+\frac{m^2}{2}\vert\phi\vert^2
 +\frac{1}{4!}u\vert\phi\vert^4,\label{XY}
\end{align}
where $\phi$ is the complex fluxon field.  The two phases of this
field theory are: 1) the fluxon uncondensed phase where
$\langle\phi\rangle=0$ and $R_{S_z}$ is unbroken. In this phase,
creating a spin fluxon costs a finite energy. In the gauge theory
jargon the $Z_2$ gauge field is in the deconfined phase. This is the
phase of a {\it spin liquid with a finite gap for spinon (bosonic)
excitations.} 2) The fluxon condensed phase where
$\langle\phi\rangle\ne 0$ and $R_{S_z}$ is spontaneously broken.
This is a phase where the $Z_2$ gauge field fluctuates so strongly
that it confines the fermionic charge excitations. Magnetically it
is an XY ordered ferromagnet. (We have implicitly assumed that the
ordering is easy plane rather than easy axis, which is natural in
the presence of spin-orbit coupling \cite{to_appear}) Moreover,
since the fermionic charge excitation are absent at low energies
throughout the transition, this phase is an electric insulator. Thus
spin fluxon condensation triggers a spin liquid to a ferromagnetic
insulator transition. According to \Eq{XY}, the universality class
of the transition is 3D XY. The fact that $\phi$ transforms as
$e^{i\theta/2}$ while the order parameter $S^+$ transform like
$e^{i\theta}$ under $R_{S_z}$ implies the identification  $S^{+}
\sim \phi^2$. Hence, there is a subtle difference from the regular
XY transition obtained from magnon condensation (i.e. condensing
$S^+$ itself), in that  the order parameter's critical scaling
dimension is anomalously large \cite{SachdevSenthil}.

(II) $S_z$ is not conserved, but {\bf T} is preserved. Now, one can
add spin rotation breaking terms to the effective Lagrangian as long
as they preserve time reversal symmetry.  The first such term in the
long wavelength limit is $g(S^+)^2+{\rm h.c.}$, is actually a
quartic term when written in terms of the spinon fields introduced
above $g\phi^4 + {\rm h.c.}$. Now, the condensation of $\phi$ leads
to a confined insulator with the spontaneous breaking of time
reversal symmetry. Interestingly, although such an insulator has an
Ising order parameter, the transition is expected to remain 3D XY
like, due to the irrelevance of four fold anisotropy at the XY
critical point.

In the past, the transition to magnetically ordered states from spin
liquids has been described using the Higgs mechanism. Here, we have
described how {\em confinement} can also lead to magnetic order.
This mechanism can lead to novel quantum phase transitions
complementing those discussed in \cite{dQCP}, which will be
described in future work \cite{to_appear}.

\acknowledgements After completing this work, we learnt that in a
recent preprint arXiv:08010252 X-L Qi and S-C Zhang have obtained
similar results\cite{qi_zhang}. We thank Joel Moore and Cenke Xu for
helpful discussions. The authors were supported by the Directior,
Office of Science, Office of Basic Energy Sciences, Materials
Sciences and Engineering Division, of the U.S. Department of Energy
under Contract No. DE-AC02-05CH11231.


\begin{thebibliography}{99}

\bibitem{tkkn} D. J. Thouless \textit{et al}, Phys. Rev. Lett. \textbf{49},
405-408 (1982).
\bibitem{haldane} F. D. M. Haldane, Phys. Rev. Lett. \textbf{61}, 2015
(1988).
\bibitem{km}C. L. Kane and E. J. Mele, Phys. Rev. Lett. \textbf{95},
226801 (2005).
\bibitem{Fu_Kane_Mele} L. Fu, C. L. Kane and E. J. Mele, Phys. Rev. Lett. \textbf{98}, 106803 (2007)
\bibitem{moore} J. E. Moore and  L. Balents, Phys. Rev. B \textbf{75}, 121306(R)
(2007).
\bibitem{spinhall} S. Murakami, N. Nagaosa and S.C. Zhang, Science \textbf{301},
1348 (2003).
\bibitem{wen}X.-G. Wen,
{\it Quantum Field Theory Of Many-body Systems: From The Origin Of
Sound To An Origin Of Light And Electrons} (Oxford University Press,
2004).
\bibitem{majorana} G. E. Volovik, JETP Lett. \textbf{70}, 609
(1999); N. Read and D. Green, Phys. Rev. B \textbf{61}, 10267
(2000); M. Stone and R. Roy, Phys. Rev. B \textbf{69}, 184511
(2004). S. Tewari, S. Das Sarma and D.-H. Lee, Phys. Rev. Lett.
\textbf{99}, 037001 (2007).
\bibitem{bernevig} B. A. Bernevig, L. T. L. Hughes and S.-C. Zhang, Science {\bf 314}, 1757 (2006);
\bibitem{lzx} D.-H. Lee, Q.-M. Zhang and T.Xiang, Phys. Rev. Lett.
in  press.
\bibitem{franz} C. Weeks {\it et al}, Nature Physics, 3 , 796 (2007). cond-mat/0703001.
\bibitem{note} Assuming initially we fix the gauge such that the two fluxons
are connected by a $\tau^z=-1$ string and $\tau^z=1$ everywhere
else. After hopping one fluxon  around a closed loop, the final
state has extra $\tau^z=-1$ on all the bonds in the whole loop. This
final state differs from the initial state by a $Z_2$ gauge
transformation and thus are the same state. \emph{This identification is essential}
for obtaining the right result.
\bibitem{SachdevSenthil} A. V. Chubukov and T. Senthil, and S.
Sachdev, Phys. Rev. Lett. 72, 2089 (1994).
%\bibitem{fisherlee} M. P.A. Fisher and D-H Lee, Phys. Rev. B
%{\bf 39}, 2756 (1989).
\bibitem{dQCP}T. Senthil {\it et al}, Science {\bf 303}, 1490
(2004).
%\bibitem{burkov}A.A. Burkov and L. Balents Physical Review B 72, 134502
%(2005).
\bibitem{to_appear} Ying Ran, Dung-Hai Lee and Ashvin Vishwanath, in
preparation.
\bibitem{qi_zhang}Xiao-Liang Qi and Shou-Cheng Zhang arXiv:08010252
\end{thebibliography}
\end{document}